\documentclass[fleqn,usenatbib,letters]{mnras}
\usepackage{newtxtext,newtxmath}
\usepackage[T1]{fontenc}
\usepackage{ae,aecompl}

\usepackage{hyperref,breakurl}

\usepackage{amsmath,amssymb,latexsym,times}

\usepackage{etoolbox}
\makeatletter
\patchcmd\@combinedblfloats{\box\@outputbox}{\unvbox\@outputbox}{}{\errmessage{\noexpand patch failed}}
\makeatother

\usepackage{graphicx}	
\usepackage{amsmath}	
\usepackage{amssymb}	
\usepackage{multirow}

\newcommand{\cosmolike}{\textsc{CosmoLike}}
\newcommand{\cosmosis}{\textsc{CosmoSIS}}

\usepackage{eso-pic}

\AddToShipoutPictureBG*{%
  \AtPageUpperLeft{%
    \hspace{0.75\paperwidth}%
    \raisebox{-3.5\baselineskip}{%
      \makebox[0pt][l]{\textnormal{DES 2018-0350}}
}}}%

\AddToShipoutPictureBG*{%
  \AtPageUpperLeft{%
    \hspace{0.75\paperwidth}%
    \raisebox{-4.5\baselineskip}{%
      \makebox[0pt][l]{\textnormal{FERMILAB-PUB-18-128-A-PPD}}
}}}%

\title[Survey geometry and cosmic shear measurements]{Survey geometry and the internal consistency of recent cosmic shear measurements}

\author[Troxel, Krause, et al. (DES Collaboration)]{
\parbox{\textwidth}{
\Large
M.~A.~Troxel,$^{1,2}$\thanks{E-mail: troxel.18@osu.edu}
E.~Krause,$^{3,4}$\thanks{E-mail: ekrause@caltech.edu}
C.~Chang,$^{5}$
T.~F.~Eifler,$^{3,4}$
O.~Friedrich,$^{6,7}$
D.~Gruen,$^{8,9}$
N.~MacCrann,$^{1,2}$
A.~Chen,$^{10}$
C.~Davis,$^{8}$
J.~DeRose,$^{11,8}$
S.~Dodelson,$^{12}$
M.~Gatti,$^{13}$
B.~Hoyle,$^{6,7}$
D.~Huterer,$^{10}$
M.~Jarvis,$^{14}$
F.~Lacasa,$^{15,16}$
P.~Lemos,$^{17,18}$
H.~V.~Peiris,$^{19}$
J.~Prat,$^{13}$
S.~Samuroff,$^{12}$
C.~S{\'a}nchez,$^{14,13}$
E.~Sheldon,$^{20}$
P.~Vielzeuf,$^{13}$
M.~Wang,$^{21}$
J.~Zuntz,$^{22}$
O.~Lahav,$^{19}$
F.~B.~Abdalla,$^{19,23}$
S.~Allam,$^{21}$
J.~Annis,$^{21}$
S.~Avila,$^{24}$
E.~Bertin,$^{25,26}$
D.~Brooks,$^{19}$
D.~L.~Burke,$^{8,9}$
A.~Carnero~Rosell,$^{16,27}$
M.~Carrasco~Kind,$^{28,29}$
J.~Carretero,$^{13}$
M.~Crocce,$^{30,31}$
C.~E.~Cunha,$^{8}$
C.~B.~D'Andrea,$^{14}$
L.~N.~da Costa,$^{16,27}$
J.~De~Vicente,$^{32}$
H.~T.~Diehl,$^{21}$
P.~Doel,$^{19}$
A.~E.~Evrard,$^{33,10}$
B.~Flaugher,$^{21}$
P.~Fosalba,$^{30,31}$
J.~Frieman,$^{21,5}$
J.~Garc\'ia-Bellido,$^{34}$
E.~Gaztanaga,$^{30,31}$
D.~W.~Gerdes,$^{33,10}$
R.~A.~Gruendl,$^{28,29}$
J.~Gschwend,$^{16,27}$
G.~Gutierrez,$^{21}$
W.~G.~Hartley,$^{19,35}$
D.~L.~Hollowood,$^{36}$
K.~Honscheid,$^{1,2}$
D.~J.~James,$^{37}$
D.~Kirk,$^{19}$
K.~Kuehn,$^{38}$
N.~Kuropatkin,$^{21}$
T.~S.~Li,$^{21,5}$
M.~Lima,$^{39,16}$
M.~March,$^{14}$
F.~Menanteau,$^{28,29}$
R.~Miquel,$^{40,13}$
J.~J.~Mohr,$^{41,42,6}$
R.~L.~C.~Ogando,$^{16,27}$
A.~A.~Plazas,$^{4}$
A.~Roodman,$^{8,9}$
E.~Sanchez,$^{32}$
V.~Scarpine,$^{21}$
R.~Schindler,$^{9}$
I.~Sevilla-Noarbe,$^{32}$
M.~Smith,$^{43}$
M.~Soares-Santos,$^{44}$
F.~Sobreira,$^{45,16}$
E.~Suchyta,$^{46}$
M.~E.~C.~Swanson,$^{29}$
D.~Thomas,$^{24}$
A.~R.~Walker,$^{47}$
R.~H.~Wechsler$^{11,8,9}$
\begin{center} (DES Collaboration) \end{center}
}
\\
(Affiliations are listed in Appendix A)
}

\date{Accepted XXX. Received YYY; in original form ZZZ}
\pubyear{2018}

\begin{document}
\label{firstpage}
\pagerange{\pageref{firstpage}--\pageref{lastpage}}
\maketitle

\begin{abstract}
We explore the impact of an update to the typical approximation for the shape noise term in the analytic covariance matrix for cosmic shear experiments that assumes the absence of survey boundary and mask effects. 
We present an exact expression for the number of galaxy pairs in this term based on the survey mask, which leads to more than a factor of three increase in the shape noise on the largest measured scales for the Kilo-Degree Survey (KIDS-450) real-space cosmic shear data. 
We compare the result of this analytic expression to several alternative methods for measuring the shape noise from the data and find excellent agreement.
This update to the covariance resolves any internal model tension evidenced by the previously large cosmological best-fit $\chi^2$ for the KiDS-450 cosmic shear data. 
The best-fit $\chi^2$ is reduced from 161 to 121 for 118 degrees of freedom. 
We also apply a correction to how the multiplicative shear calibration uncertainty is included in the covariance. 
This change shifts the inferred amplitude of the correlation function to higher values.
We find that this improves agreement of the KiDS-450 cosmic shear results with Dark Energy Survey Year 1 and \textit{Planck} results. 
\end{abstract}


\section{Introduction}

Cosmic shear, the weak lensing of background galaxies by large-scale structure in the Universe, has been viewed for many years as a promising cosmological probe \citep{detf,esoesa,weinberg13}. With the current generation of surveys, among them the Kilo-Degree Survey (KiDS, \citealt{kids2015})\footnote{http://kids.strw.leidenuniv.nl} and the Dark Energy Survey (DES),\footnote{http://www.darkenergysurvey.org} we are beginning to see this promise fulfilled. Alongside the improved statistical power of these latest data sets, weak lensing methodology has developed rapidly over the last several years \citep{2017MNRAS.467.1627F,HuffMandelbaum2017,SheldonHuff2017,shearcat,2017arXiv171000885M,2018PASJ...70S..25M}. With DES and KiDS, we have two independent weak lensing data sets of comparable constraining power -- capable of constraining the amplitude of the two-point correlation function at better than the 3\% level in the case of DES. If the reduction of systematic uncertainties can keep pace, these constraints will rapidly improve over the coming years as the area of sky analysed and the integrated exposure time of each survey increases by factors of several. A third, similarly powerful constraint from the Hyper-Suprime Cam (HSC) survey\footnote{http://hsc.mtk.nao.ac.jp/ssp/} is also expected on a similar timescale. The Canada-France-Hawaii Telescope Lens Survey \citep{joudaki17} and the Deep Lens Survey \citep{jee2015} have also published cosmic shear constraints of similar precision to those of KiDS-450.

Cosmic shear is one of several low-redshift probes that tightly constrain the amplitude of structure in the standard cosmological model ($\Lambda$CDM). 
By comparing these constraints to high-redshift measurements like the cosmic microwave background (CMB), we are able to fundamentally test the ability of $\Lambda$CDM to predict observations of the structure in and geometry of the Universe across its entire history.
One important question is whether the amplitude of structure measured by lensing surveys is compatible with predictions based on CMB observations, extrapolated to late times assuming $\Lambda$CDM. In order to interpret these results as strong, reliable tests of $\Lambda$CDM, we must first subject them to stringent tests of \emph{internal} and \emph{mutual} consistency. 

\emph{Internal} consistency tests the agreement of different subsets of the weak lensing data vector with predictions based on the same $\Lambda$CDM parameters. The most simple such test is whether the best-fitting $\Lambda$CDM model describes these measurements (i.e., the correlation function) at an acceptable goodness-of-fit ($\chi^2$). Both the latest DES \citep[][T17]{shearcorr} and KiDS \citep[][H17]{kids450} cosmic shear analyses reported a large $\chi^2$ per degree of freedom (dof) in their initially released results.

Tests of \emph{mutual} consistency, e.g.~between the fully independent cosmic shear real-space two-point correlation function from KiDS-450 (H17) and DES Y1 (T17) are a unique opportunity to validate the entire analysis process. We note that mutual consistency between the DES and KiDS cosmic shear results is not a new result in this work -- good agreement between the two data sets has already been demonstrated in T17.

In this paper we make substantial progress on these critical consistency tests by implementing three improvements:\footnote{We note that these effects were known to the KiDS collaboration before the publication of this paper, with updates already incorporated for their upcoming 9-band cosmological analysis (Hildebrandt et al in prep., KiDS Collaboration private communication).} 
\begin{enumerate}
\item to the shape noise component of analytic cosmic shear covariance matrix estimates relative to the initial implementation in both analyses. This correction was first implemented during the referee process for the DES Y1 analyses (\cite{methodpaper}, K17; \cite{keypaper}; T17). 
We find that it non-trivially impacts current generation weak lensing surveys, especially in the regime of small fields or non-compact\footnote{By compact, we simply mean regions that minimize boundary length relative to area, i.e., a circle.} footprints, including the KiDS-450 covariance matrix (H17) described below;
\item to the treatment of uncertainty in the multiplicative shear bias calibration in the H17 covariance matrix; and
\item to the angular separation estimate used for evaluating the predicted shear signal for the bins in H17 (see footnote 1 of \citealt{2017arXiv170706627J}). 
\end{enumerate}
The first update has a significant impact on the interpretation of internal consistency in both H17 and T17, particularly the best-fit $\chi^2$. The shape noise component of the covariance in such surveys should thus be treated with care beyond the simple geometric approximations commonly used in earlier work. Each of the latter two updates shift the H17 contours only by a fraction of the 68\% CL, but jointly are relevant for evaluating mutual consistency and consistency with CMB observations.

This paper is organised as follows: In Sec.~\ref{sec:cs}, we describe the cosmic shear data used and our analysis framework relative to H17 and T17. 
We describe methods to appropriately describe the shape noise term in non-compact or disjoint survey geometries and discuss how this impacts the H17 results in Sec.~\ref{sec:cov}, including also update (ii) mentioned above. 
We then discuss the resulting interpretation of the cosmic shear results from H17 and T17 in Sec.~\ref{sec:cosmo}. 
We conclude and discuss the outlook for future such studies in Sec.~\ref{sec:conc}.

\section{Cosmic Shear Data \& Analysis}\label{sec:cs}

We infer cosmological model constraints from the KiDS-450 and DES Y1 datasets using measurements of the cosmic shear real-space correlation function between tomographic bins $i$ and $j$,
\begin{equation}
\xi^{ij}_{\pm}(\theta) = \frac{1}{2\pi}\int d\ell \ell J_{0/4}(\theta \ell) P^{ij}_{\kappa}(\ell) \; .
\label{eq:Pkappa}
\end{equation}
The corresponding convergence power spectrum is $P^{ij}_{\kappa}$, and $J_{\alpha}$ is a Bessel function of the first kind. $P_{\kappa}$ is related to the nonlinear matter power spectrum $P_{\textrm{NL}}$ via the Limber approximation as
\begin{equation}
P^{ij}_{\kappa}(\ell) = \int_0^{\chi_H}d\chi \frac{q^i(\chi)q^j(\chi)}{\chi^2} P_{\textrm{NL}}\left(\frac{\ell+1/2}{\chi},\chi\right).
\label{eq:Pdelta}
\end{equation}
Here $\chi$ is the radial comoving distance and $\chi_H$ is the horizon distance. The lensing efficiency function $q$,
\begin{equation}
q^i(\chi) = \frac{3}{2}\Omega_m \left(\frac{H_0}{c}\right)^2 \frac{\chi}{a(\chi)} \int_{\chi}^{\chi_H} d\chi' n^i(\chi')\frac{\chi'-\chi}{\chi'} \; ,
\end{equation}
depends on the matter density $\Omega_m$, Hubble constant $H_0$, scale factor $a$, and the effective number density distribution of galaxies $n^i(\chi)$, normalised such that $\int d\chi n^i(\chi)=1$. 

We assume a multivariate Gaussian in the correlation function measurements
\begin{equation}
\ln\mathcal{L}(\mathbf{p}) = -\frac{1}{2}\sum_{ij} [D_i-T_i(\mathbf{ p})]\, C^{-1}{}_{ij}\, [D_j-T_j(\mathbf{p})] \; ,
\end{equation}
where $\mathbf{p}$ is the set of parameters and $T_i(\mathbf{ p})$ are the theoretical predictions for $\xi_{\pm}$. We utilise the data vector $\xi^{ij}_{\pm}$ ($D$), redshift distributions, covariance matrix, scale cuts, and priors from the cosmic shear analyses of both H17 and T17, wherein further modelling details can be found. 
In these works, the analysis of the cosmic shear signal is distinct in almost every step of the shape measurement and subsequent modelling process. 
This includes the shape measurement algorithms, shear calibration methods, redshift calibration strategy, theory modelling choices, and cosmological model parameterisation. The KiDS data\footnote{KiDS is an ESO public survey.  All data products are available to download from http://kids.strw.leidenuniv.nl/DR3/index.php.} are processed by THELI \citep{2013MNRAS.433.2545E} and Astro-WISE \citep{2013ExA....35....1B,2015A&A...582A..62D}. Shapes are measured using \textsc{lensfit} \citep{2013MNRAS.429.2858M}, and photometric redshifts are obtained from PSF-matched photometry and calibrated using external overlapping spectroscopic surveys \citep[see][]{kids450}. Details of the DES Y1 data processing can be found in \citealt{shearcat,y1gold} and references therein.

Because the analysis choices are fundamentally different between H17 (their Table 3) and T17 (their Table 2), we test conclusions in this work using both analysis configurations. 
These differences include parameter choices, such as sampling over $\Omega_m$ vs. $\Omega_m h^2$, as well as changes to the angular scales used and nonlinear matter power spectrum modelling. The impact of baryonic effects, which are modelled in the H17 analysis configuration, are mitigated in the T17 analysis configuration by removing additional angular scales from the data vector that may be biased by baryonic effects. 

We make two important changes to the H17 data vector and covariance: 1) we update the measured $\xi^{ij}_{\pm}$ to use the weighted mean pair separation as noted in footnote 1 of \cite{2017arXiv170706627J}, and 2) we update the covariance matrix as discussed below in Sec. \ref{sec:cov}. 
Rather than use the multiple realisations of the redshift distribution as in H17, we sample over four photometric redshift bias parameters on the mean redshift of each tomographic bin with Gaussian widths $\sigma(\Delta_z^i)\times 10^2 = \{2.376, 1.380, 0.945, 0.594\}$. The priors for these parameters are taken from the error on the mean redshift given in Appendix A3 of H17. We verify that this last change, as well as that due to sampling over $H_0$ instead of $\theta_{\mathrm{mc}}$ (100 times the ratio of the sound horizon to the angular diameter distance), results in only a slight difference from the original published H17 contour in $S_8$ and $\Omega_m$, which does not impact any conclusions in this work.\footnote{For example, we reproduce the peak 1-D value of $S_8$ in the original H17 chain to 0.03\%, while finding a 0.14$\sigma$ weaker constraint.}

The results in this work are derived using \cosmosis\footnote{https://bitbucket.org/joezuntz/cosmosis/wiki/Home} \citep{cosmosis}, which uses \textsc{CAMB} \citep{Lewis2000,howlett2012} and has been compared in detail with \cosmolike\ \citep{ke16} in K17. For the H17 analysis configuration, we model the nonlinear matter power spectrum and baryonic effects using \cite{mead,mead2}. In the T17 analysis configuration, we use \textsc{HALOFIT} \citep{SPJ+03} with updates from \cite{takahashi2012}. The impact of neutrino mass on the matter power spectrum is implemented in \textsc{HALOFIT} from \cite{nu}. We sample the parameter spaces, with dimensionality ranging from 11 to 16, using $\textsc{MultiNest}$ \citep{mn3}.

\section{Modelling the Covariance}\label{sec:cov}
The fiducial analyses of H17 and T17 both use covariances obtained from analytic models (see H17, K17 for model details and validation). 
These analytic covariances consist of a Gaussian part, which combines cosmic variance and shape noise, and a non-Gaussian part, which includes contributions from the connected matter trispectrum as well as super-sample covariance \citep{TakadaHu13}.

The pure shape noise contribution to the covariance $\mathrm{Cov^{SN}}$ is given by \citep{Schneider02b,Joachimi08}
\begin{equation}
\mathrm{Cov^{SN}}\left(\xi^{ij}_\pm(\theta_1),\xi^{kl}_\pm(\theta_2)\right)  = \frac{(\sigma_\epsilon^i \sigma_\epsilon^j)^2}{N^{ij}_\mathrm{p}(\theta_1,\Delta \theta_1)}\delta_{\theta_1\theta_2}\left(\delta_{ik}\delta_{jl}+\delta_{il}\delta_{jk}\right)\,,\label{eq:covsn}
\end{equation}
where $i$, $j$, $k$, $l$ are tomographic bin indices, $\sigma_\epsilon$ is the ellipticity dispersion, $N_\mathrm{p}(\theta,\Delta \theta)$ is the number of galaxy pairs in angular bin $\Delta \theta$, and $\delta$ is the Kronecker delta function. While \citet{Schneider02b} provide an exact expression for $N_\mathrm{p}$ as the weighted sum over all galaxy pairs, this is commonly approximated with $N^{ij}_\mathrm{p}(\theta) = 2 \pi \theta \Delta\theta A \bar{n}^i\bar{n}_{\mathrm{eff}}^j$, where $A$ is the survey area, $\bar{n}_{\mathrm{eff}}$ the mean effective projected galaxy density, and $ 2 \pi \theta \Delta\theta$ the approximate area of the annular bin. We will refer to this approximation as the geometric shape noise estimate. In all cases, we use $n_{\mathrm{eff}}$ and $\sigma_e$ as defined in \citep{heymans13}, where the total $\sigma_e$ is used.

We revisit this common geometric estimate and instead calculate $N_\mathrm{p}$ directly for an infinitesimal annular bin element, taking into account both the survey geometry, as well as the clustered distribution of source galaxies. This calculation starts from the continuous limit of the discrete pair count expression of \citet{Schneider02b},
\begin{align}
\nonumber N^{ij}_\mathrm{p}\!(\theta,d\theta) \!&= \int_{S^2}\!\!\!\!\!d^2\theta'\int_{|\boldsymbol\theta''|\in [\theta,\theta+d\theta]}\!\!\!\!\!\!\!\!\!\!\!\!\!\!\!\!\!\!\!\!\!\!\! d^2 \theta'' W(\boldsymbol\theta')n^i(\boldsymbol\theta')W(\boldsymbol\theta''\!\!\!-\boldsymbol\theta')n^j(\boldsymbol\theta''\!\!\!-\boldsymbol\theta')\\
&\nonumber =2 \pi \theta A \bar{n}^i\bar{n}^j\Big \langle W(\boldsymbol\theta')\mathfrak{d}^i(\boldsymbol\theta')\! W(\boldsymbol\theta''\!\!\!-\!\boldsymbol\theta')\mathfrak{d}^j(\boldsymbol\theta''\!\!\!-\!\boldsymbol\theta')\Big\rangle_{|\boldsymbol\theta''|\in [\theta,\theta+d\theta]} d\theta\\
&= 2 \pi \theta A \bar{n}^i\bar{n}^j\left[w_W(\theta)\left(1+w^{ij}_{\mathrm g}(\theta)\right)\right]d\theta\; .\label{eq:mask}
\end{align}
Here $W(\boldsymbol\theta)$ is the angular survey mask, normalised such that $ \int_{S^2}d^2\theta W(\boldsymbol\theta) = A$, $n^i(\boldsymbol\theta)$ is the projected galaxy density field, and $\mathfrak{d}^i(\boldsymbol\theta) = 1\!+\!\delta^i_{\mathrm g}(\boldsymbol\theta)$, where $\delta_{\mathrm g} (\boldsymbol\theta)$ is the projected source galaxy density contrast. $W$ and $\delta_{\mathrm g}$ have angular correlation functions $w_W$ and $w^{ij}_{\mathrm g}$. The latter is often called source clustering, and we define $w_W$ to be normalised such that  $w_W(0) = 1,\,\, w_W(\theta \rightarrow \infty) = 0$. We compute $w_W$ from the the power spectrum of the mask, which is evaluated using the {\sc Healpix}\footnote{\url{http://healpix.sf.net}} sphere pixelization software package \cite{2005ApJ...622..759G}.

\begin{figure}
\begin{center}
\includegraphics[width=\columnwidth]{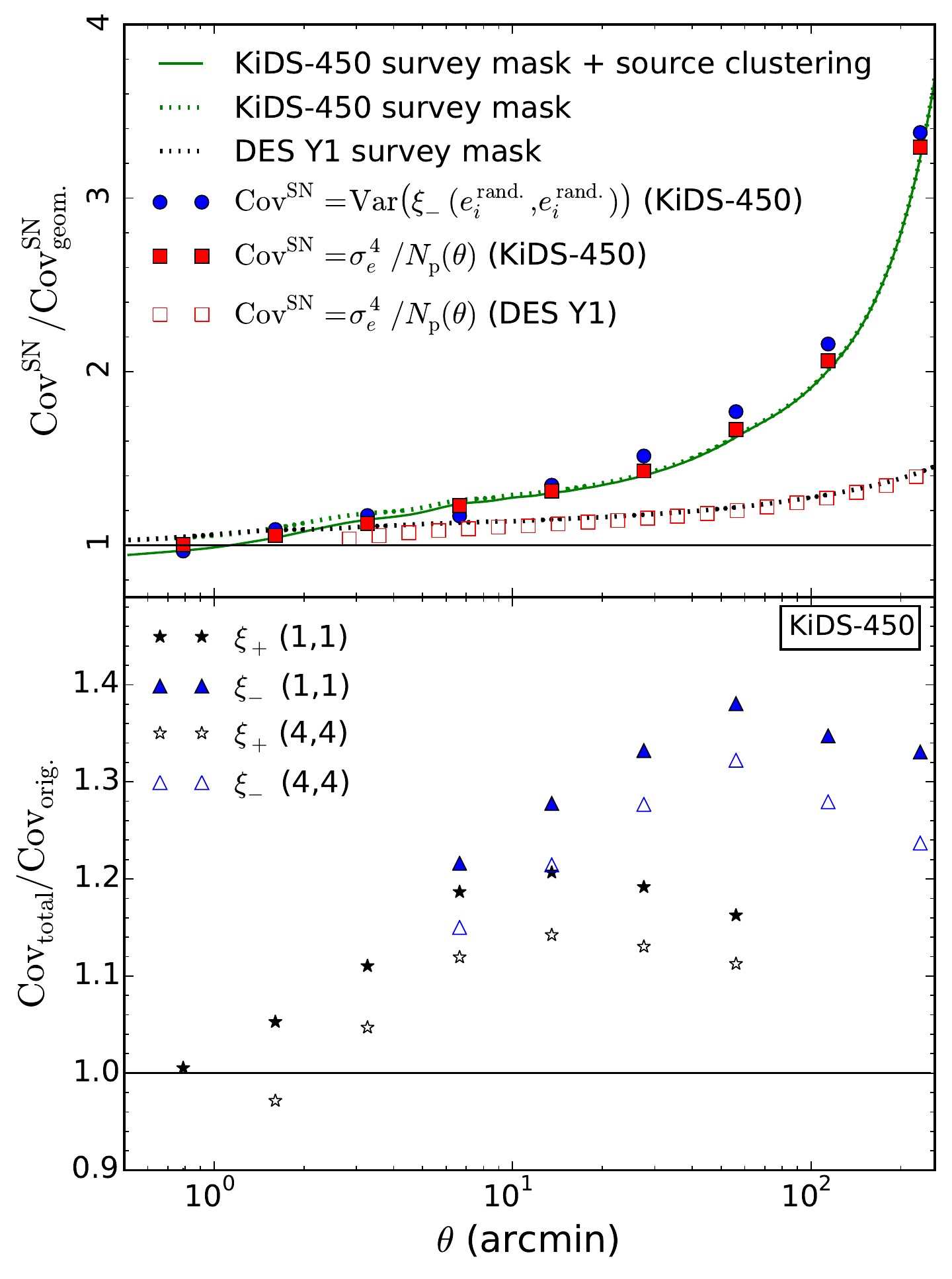}
\end{center}
\caption[]{The impact of updates to the KiDS-450 covariance diagonal.
\textit{Upper panel}: The ratio of the true shape noise term to the geometric approximation in the (1,1) tomographic bin pair, which has a nominal range of $z=0.1$ to $0.3$, from 
(a) the variance of $\xi_-$ (Var($\xi_+$) is also consistent) measured from 1000 random rotations of the KiDS-450 shape catalogue (blue circles), 
(b) replacing $n_{\mathrm{eff}}$ with the measured $N_{\mathrm{p}}(\theta)$ in Eq. \ref{eq:covsn} (red squares), and 
(c) an analytic prediction for $N_{\mathrm{p}}(\theta)$ from the survey mask in Eq. \ref{eq:mask} with (solid) and without (dotted) source clustering (green lines). 
We compare the impact for a DES Y1-like survey footprint.
For KiDS-450, the correction increases the shape noise by up to a factor of 3.5 on the largest measured scales, which corresponds to a maximum factor of 1.4 for DES Y1. 
On the smallest scales, the shape noise is slightly decreased due to source clustering. 
\textit{Lower panel}: The ratio of the final corrected covariance to the original covariance for $\xi_{+}$ (black stars) and $\xi_{-}$ (blue triangles). Only angular scales used in the H17 analysis are shown. We include the (4,4) tomographic bin pair with nominal range $z=0.7$ to $0.9$ for comparison (open symbols).
\label{fig:sn}}
\end{figure}

In addition to the analytic calculation of $N_{\mathrm{p}}$, we can also measure components of the shape noise in Eq. \ref{eq:covsn} directly from the data. One way to do this is by counting the effective number of pairs of galaxies in the survey footprint, $N_{\mathrm{p}}(\theta)$, to use in Eq. \ref{eq:covsn}. This approach still makes use of the measured $\sigma_\epsilon$. The total shape noise contribution $\mathrm{Cov^{SN}}$ can also be measured from the variance of $\xi_{\pm}$ across many random realisations of the shape catalogue \citep[e.g.,][]{2017arXiv171005045G,2017MNRAS.465..746S,2018ApJ...854..120M}, where $\xi_{\pm}$ is measured in each realisation after randomly rotating the measured galaxy shapes. 

We note that survey geometry also affects the other terms in the analytic covariance: the impact of the KiDS footprint on super-sample variance is included in the analytic covariance from H17,  and the impact of survey geometries on the cosmic variance and mixed covariance terms is explored in \citet{Kilbinger04,Joachimi08}. \cite{2017MNRAS.471.3827S} also considers the effect of masking on the covariance of several large-scale structure probes. However, in this analysis we focus on the shape noise, which is the leading contribution to the total variance on most angular scales included in the cosmology analysis.

\begin{figure*}
\begin{center}
\includegraphics[width=\textwidth]{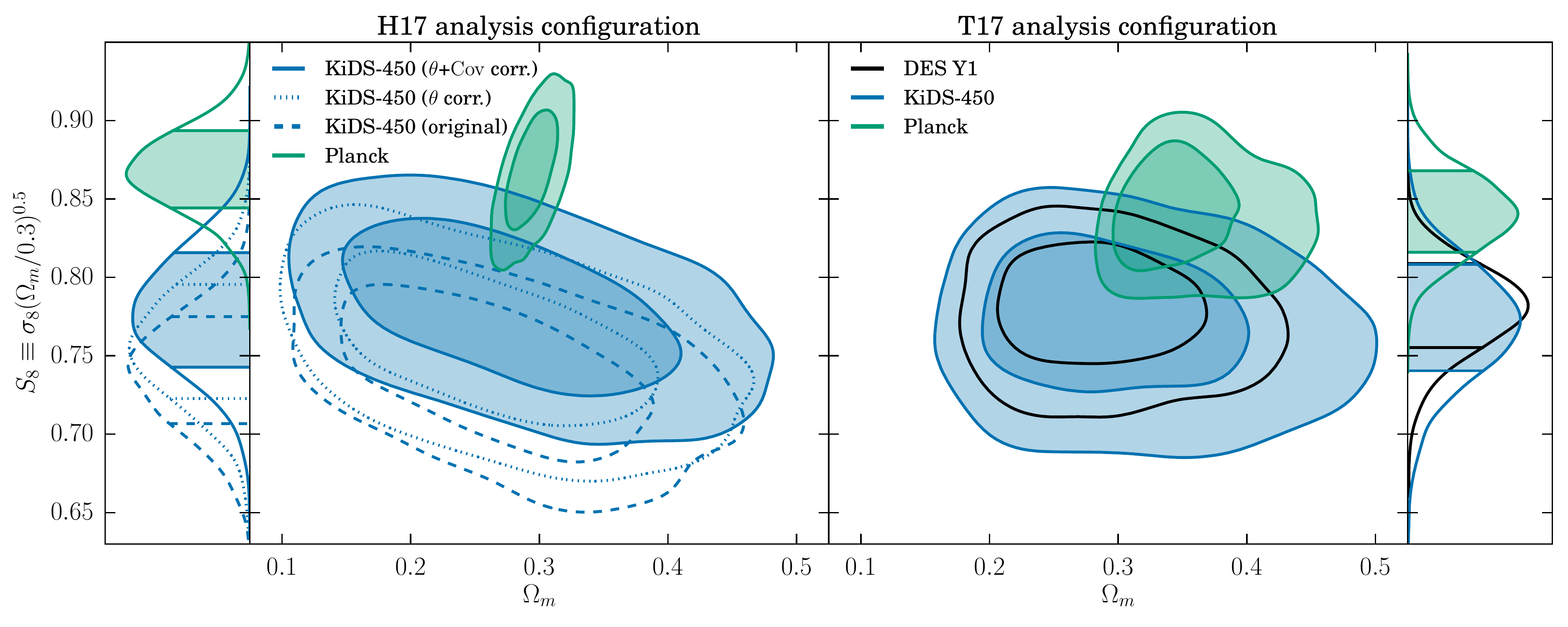}
\end{center}
\caption[]{\textit{Left panels}: The impact of data vector and covariance corrections on the KiDS-450 cosmic shear results in the H17 analysis configuration. '$\theta$ corr.' refers to the update of the $\theta$ values for the data vector that appropriately averages the mean pair separation noted in Footnote 1 of \cite{2017arXiv170706627J}. '$\theta$+$\mathrm{Cov}$ corr.' refers to additionally including the covariance corrections discussed in Sec. \ref{sec:cov} -- updating the $\mathrm{Cov}^{\mathrm{SN}}$ and $\sigma_m$ components. The $\mathrm{Cov}^{\mathrm{SN}}$ update alone has relatively little impact on the cosmological constraints compared to the $\sigma_m$ change.
\textit{Right panels}: A comparison of the final cosmic shear results from the KiDS-450 and the DES Y1 data in the T17 analysis configuration. 
In both panels, we include constraints from the CMB (\textit{Planck}) for comparison, analysed separately in the two analysis configurations, and show the marginalised $S_8$ constraints on each side.
Note that, among other differences described in the text, the neutrino mass density is fixed in the left panels (H17) and marginalized over in the right panels (T17), which causes the Planck contours in particular to differ.
The cosmic shear results of the DES and KiDS analyses are strongly consistent, though the complete overlap found here is likely coincidental and not necessarily expected statistically. The 2-D 68\% CL of both overlap with those of the CMB in the right panels (and nearly so in the left panels).
\label{fig:cosmo}}
\end{figure*}

\subsection{Impact of survey geometry on shape noise}\label{sec:sn2}

We demonstrate that the three approaches to measuring $\mathrm{Cov^{SN}}$ give consistent results for KiDS-450 in Fig. \ref{fig:sn}, while they disagree substantially with the geometric estimate.
We find that in surveys that are non-compact or composed of multiple small fields, like KiDS-450, measurements of the shape noise that account for survey geometry can be significantly boosted relative to the standard geometric shape noise estimate. 
This update slightly reduces the impact of shape noise on scales $<2$ arcmin due to source clustering, while significantly boosting the measured shape noise on large scales, up to a factor of 3.5 relative to the geometric estimate at 300 arcmin. We also show the impact of the shape noise update to the total covariance in KiDS-450 in the lower panel of Fig. \ref{fig:sn}.

This is less of an issue for survey geometries that are more contiguous and compact, such as DES Y1, but there is still a non-negligible difference in the shape noise term compared to the geometric estimate. This is shown in the top panel of Fig. \ref{fig:sn}, which compares the effect on the KiDS-450 and DES Y1 analyses. 
This difference will diminish as survey footprints become larger and for those that are designed to be contiguous and compact.

\subsection{Treatment of multiplicative shear bias uncertainty}\label{sec:sn3}

In addition to the shape noise update described in Sec. \ref{sec:sn2}, we also modify how the multiplicative shear bias uncertainty $\sigma_m$ is added to the covariance, which is given in Eq. 12 of H17: 
\begin{equation}
\mathrm{Cov}^{ij}_{\mathrm{cal}} = 4\sigma_m^2\xi_\pm^{i}\xi_\pm^{j} + \mathrm{Cov}^{ij}.\label{eq:sigmam}
\end{equation}
This approach of analytic marginalisation should be equivalent for a single $m$ value to what was done in T17 \citep[e.g.,][]{2002MNRAS.335.1193B} when the predicted (theory) template ($\xi_\pm^{i}$) is used. 

In the original H17 covariance, the measured $\xi_{\pm,\mathrm{meas.}}$ was used as a fixed template to evaluate this term, which can lead to a bias in the inferred amplitude of the signal (e.g., $S_8$) due in part to there being a larger relative addition to the covariance for the data points which scatter high due to noise than for those that scatter low.
The use of the data vector as a template in Eq. \ref{eq:sigmam} was identified by accident while separating terms in the H17 covariance to update the shape noise component described in the previous section. 
We agreed that this should be changed if it non-negligibly biased the resulting cosmological inference prior to knowing its precise impact on the inferred parameter values. 

After we decided to change the template used in Eq. \ref{eq:sigmam}, we further explored the impact of using various fixed templates for the analytic marginalisation of $m$ relative to the formally correct, cosmology-dependent template in Eq. \ref{eq:sigmam}. 
This included 1) using a fixed predicted $\xi_{\pm}$ as the template at the cosmology used to evaluate the other covariance terms, 2) using fixed predicted templates at two cosmologies with amplitudes of $\pm10\%$, 3) marginalising over m by including it as an additional model parameter in the inference, and 4) using the data vector as a fixed template.
The first three approaches were found to give consistent results to the cosmology-dependent approach, while using the data as a fixed template biases the inferred $S_8$ low, as shown in Fig. \ref{fig:cosmo}. 
For consistency, we implement Eq. \ref{eq:sigmam} using a fixed template at the same cosmology used to evaluate the rest of the covariance in the results described below.

\section{Interpretation of Cosmic Shear Results}\label{sec:cosmo}

The accuracy of the cosmic shear covariance can significantly impact the interpretation of measurements. 
We have found in the recent DES Y1 analyses that underestimating the shape noise by 10-30\% on most scales due to the effects described in Sec. \ref{sec:cov} can significantly worsen the best-fit reduced $\chi^2$ for a cosmological model, even if it does not (as found in T17) significantly modify the resulting cosmological constraints. 
The effect is larger for the KiDS-450 geometry, due to the presence of multiple disjoint fields, leading to changes in both the $\chi^2$ and, to a small degree, the inferred model parameters. 
In the original parameter space of H17, we find significant improvements to the $\Lambda$CDM best-fit $\chi^2$ of the KiDS-450 cosmic shear data due to the shape noise update. 
The best-fit $\chi^2$ is reduced from 161 to 121 for 118 dof, corresponding to an increase in the $p$-value from $5\times 10^{-3}$ to $0.4$.
Similarly, in the parameter space of T17, the best-fit $\chi^2$ is reduced from 122 to 78 for 67 dof. 
The interpretation of the DES cosmic shear best-fit $\chi^2$ in this parameter space is similar, with a $\chi^2$ that was reduced from 268 to 227 for 211 dof, with $p=0.21$.

The update to the way $\sigma_m$ is included in the covariance in Eq. \ref{eq:sigmam} more strongly impacts the inferred cosmology, while not significantly modifying the $\chi^2$. This is shown in the left panels of Fig. \ref{fig:cosmo} combined with the shape noise update (solid contour), along with the impact of updating the reported $\theta$ value in the data vector (dotted), both relative to the analysis of the original data vector and covariance (dashed). Both of these updates increase the inferred $S_8$.
We find a similar shift in $S_8$ in the T17 parameter space. 
This shift improves agreement in the $S_8$--$\Omega_m$ plane compared to both DES Y1 (T17) and \textit{Planck} \citep[TT+lowP]{Ade:2015xua} results. However, the complete overlap of the KiDS and DES constraints found here is likely coincidental and not necessarily expected statistically.

We compare the final parameter constraints from KiDS-450 and DES Y1 in the right panels of Fig \ref{fig:cosmo}, finding complete overlap of the KiDS-450 and DES Y1 cosmic shear contours in $S_8$ and $\Omega_m$, with constraints of $S_8=0.782^{+0.027}_{-0.027}$ for DES Y1 and $S_8=0.772^{+0.037}_{-0.031}$ for KiDS-450 in the T17 analysis configuration.
Beyond the primary cosmological parameters, it is also important to recognise \citep[as recently highlighted in][]{2018MNRAS.476..151E} the impact that the major astrophysical systematic in cosmic shear, the intrinsic alignment of galaxies (IA) \citep[see][and references therein]{Joachimi2015,Troxel20151}, can have on the interpretation of cosmological results. 
One diagnostic of potential residual systematics is an inconsistent model fit for the IA signal, up to a potential difference in the effective amplitude due to the use of different shape measurement methods. 
We also find excellent agreement here, with an amplitude for the intrinsic alignment model of $A_{\mathrm{IA}}=1.0^{+0.4}_{-0.7}$ (DES Y1) and $A_{\mathrm{IA}}=0.9^{+0.9}_{-0.6}$ (KiDS-450) in the T17 analysis configuration, marginalising over a free redshift power-law evolution which is also strongly consistent.
This is a powerful demonstration of consistency between the cosmic shear analyses of these two surveys, which lends credence to the robustness of constraints shown here from cosmic shear.

\section{Conclusions \& Outlook}\label{sec:conc}

We have demonstrated that using an exact measurement (e.g., the actual $N_{\mathrm{p}}(\theta)$) of the shape noise component of analytic cosmic shear covariance matrix estimates is critical for ongoing and future analyses where the survey footprint is non-compact or disjoint. In the case of KiDS-450, we have demonstrated that this correction increases the shape noise term in the covariance by up to a factor of 3.5 on the largest scales.
This shape noise correction is sufficient to completely resolve the large best-fit reduced $\chi^2$ for $\Lambda$CDM from the original analysis of H17, and the first pre-print version of T17. With these updates, there is no longer any evidence for a lack of internal model consistency in this basic test for these cosmic shear analyses. We find that these changes can also relieve previously discussed tensions in other internal consistency tests, such as those performed in \citet{2018MNRAS.476..151E}. For the six subsets of the data considered in that work, we find that all subsets are now consistent using the statistical methods described therein.

We find that two additional updates in (1) the addition of $\sigma_m$ to the covariance matrix described in Sec. \ref{sec:sn3} and (2) the determination of the effective angular values for the data vector both shift the inferred $S_8$ from KiDS-450 to slightly larger values. This improves the mutual consistency in cosmological constraints between the KiDS-450 and DES Y1 cosmic shear data sets found in T17, while also bringing the KiDS-450 and \textit{Planck} results into better agreement in the $S_8$--$\Omega_m$ plane. These results are an important step forward in the mutual validation of cosmic shear results. A more complete comparison of the DES and KiDS weak lensing results and a full investigation of the impact of survey geometry on the mixed and cosmic variance covariance terms is warranted and is left to future work. An extended study of the internal and mutual consistency between several existing weak lensing surveys, including KiDS-450, will be presented in Chang et al. in prep.

Our results weaken evidence that $\Lambda$CDM can not consistently describe both low-redshift cosmic shear and the CMB, given the agreement shown here between DES Y1, KiDS-450, and \textit{Planck}. 
With the next releases of DES, HSC, and KiDS weak lensing results and CMB results from \textit{Planck}, $\Lambda$CDM will face a much stronger test. 
These upcoming results will determine whether the current agreement converges further, or whether we begin to see evidence of new fundamental physics needed to describe the evolution of the Universe from the surface of last scattering to the low redshifts probed by weak lensing. 

\section*{Acknowledgements}

We thank O.~Dor\'e, G.~Efstathiou, C.~Heymans, H.~Hildebrandt, C.~Hirata, B.~Joachimi, S.~Joudaki, and K.~Kuijken for comments, and the KiDS Collaboration for providing their weak lensing data to the community. Part of this research was supported from NASA grant 15-WFIRST15-0008 Cosmology with the High Latitude Survey WFIRST Science Investigation Team. Part of the research was carried out at the Jet Propulsion Laboratory, California Institute of Technology, under a contract with the National Aeronautics and Space Administration and is supported by NASA ROSES ATP 16-ATP16-0084 grant and by NASA ROSES 16-ADAP16-0116. PL acknowledges support from an Isaac Newton Studentship at the University of Cambridge and from the Science and Technologies Facilities Council.

EK thanks the Centro de Ciencias Pedro Pascual in Benasque, Spain for hospitality during the workshop Understanding cosmological observations, where some of this work originated. Funding for the DES Projects has been provided by the DOE and NSF(USA), MEC/MICINN/MINECO(Spain), STFC(UK), HEFCE(UK). NCSA(UIUC), KICP(U. Chicago), CCAPP(Ohio State), MIFPA(Texas A\&M), CNPQ, FAPERJ, FINEP(Brazil), DFG(Germany) and the Collaborating Institutions in the Dark Energy Survey.

The Collaborating Institutions are Argonne Lab, UC Santa Cruz, University of Cambridge, CIEMAT-Madrid, University of Chicago, University College London, 
DES-Brazil Consortium, University of Edinburgh, ETH Z{\"u}rich, Fermilab, University of Illinois, ICE (IEEC-CSIC), IFAE Barcelona, Lawrence Berkeley Lab, 
LMU M{\"u}nchen and the associated Excellence Cluster Universe, University of Michigan, NOAO, University of Nottingham, Ohio State University, University of 
Pennsylvania, University of Portsmouth, SLAC National Lab, Stanford University, University of Sussex, Texas A\&M University, and the OzDES Membership Consortium.

This work is based in part on observations at Cerro Tololo Inter-American Observatory, National Optical Astronomy Observatory, which is operated by the Association of 
Universities for Research in Astronomy (AURA) under a cooperative agreement with the National Science Foundation.

The DES Data Management System is supported by the NSF under Grant Numbers AST-1138766 and AST-1536171. 
The DES participants from Spanish institutions are partially supported by MINECO under grants AYA2015-71825, ESP2015-66861, FPA2015-68048, SEV-2016-0588, SEV-2016-0597, and MDM-2015-0509, 
some of which include ERDF funds from the European Union. IFAE is partially funded by the CERCA program of the Generalitat de Catalunya.
Research leading to these results has received funding from the European Research
Council under the European Union's Seventh Framework Program (FP7/2007-2013) including ERC grant agreements 240672, 291329, and 306478.
We  acknowledge support from the Australian Research Council Centre of Excellence for All-sky Astrophysics (CAASTRO), through project number CE110001020, and the Brazilian Instituto Nacional de Ci\^enciae Tecnologia (INCT) e-Universe (CNPq grant 465376/2014-2).

This manuscript has been authored by Fermi Research Alliance, LLC under Contract No. DE-AC02-07CH11359 with the U.S. Department of Energy, Office of Science, Office of High Energy Physics. The United States Government retains and the publisher, by accepting the article for publication, acknowledges that the United States Government retains a non-exclusive, paid-up, irrevocable, world-wide license to publish or reproduce the published form of this manuscript, or allow others to do so, for United States Government purposes.

Based on data products from observations made with ESO Telescopes at the La Silla Paranal Observatory under programme IDs 177.A-3016, 177.A-3017 and 177.A-3018.

This research used resources of the National Energy Research Scientific Computing Center, a DOE Office of Science User Facility supported by the Office of Science of the U.S. Department of Energy under Contract No. DE-AC02-05CH11231. This work also used resources at the Ohio Supercomputing Center \citep{OhioSupercomputerCenter1987}
Some of the results in this paper have been derived using the HEALPix \cite{2005ApJ...622..759G} package.
\bibliographystyle{mnras}
\bibliography{y1a1_cosmic_shear,../../des_bibtex/des_y1kp_short}

\appendix
\section{Affiliations}
$^{1}$ Center for Cosmology and Astro-Particle Physics, The Ohio State University, Columbus, OH 43210, USA\\
$^{2}$ Department of Physics, The Ohio State University, Columbus, OH 43210, USA\\
$^{3}$ Department of Astronomy/Steward Observatory, 933 North Cherry Avenue, Tucson, AZ 85721-0065, USA\\
$^{4}$ Jet Propulsion Laboratory, California Institute of Technology, 4800 Oak Grove Dr., Pasadena, CA 91109, USA\\
$^{5}$ Kavli Institute for Cosmological Physics, University of Chicago, Chicago, IL 60637, USA\\
$^{6}$ Max Planck Institute for Extraterrestrial Physics, Giessenbachstrasse, 85748 Garching, Germany\\
$^{7}$ Universit\"ats-Sternwarte, Fakult\"at f\"ur Physik, Ludwig-Maximilians Universit\"at M\"unchen, Scheinerstr. 1, 81679 M\"unchen, Germany\\
$^{8}$ Kavli Institute for Particle Astrophysics \& Cosmology, P. O. Box 2450, Stanford University, Stanford, CA 94305, USA\\
$^{9}$ SLAC National Accelerator Laboratory, Menlo Park, CA 94025, USA\\
$^{10}$ Department of Physics, University of Michigan, Ann Arbor, MI 48109, USA\\
$^{11}$ Department of Physics, Stanford University, 382 Via Pueblo Mall, Stanford, CA 94305, USA\\
$^{12}$ Department of Physics, Carnegie Mellon University, Pittsburgh, Pennsylvania 15312, USA\\
$^{13}$ Institut de F\'{\i}sica d'Altes Energies (IFAE), The Barcelona Institute of Science and Technology, Campus UAB, 08193 Bellaterra (Barcelona) Spain\\
$^{14}$ Department of Physics and Astronomy, University of Pennsylvania, Philadelphia, PA 19104, USA\\
$^{15}$ D\'{e}partement de Physique Th\'{e}orique and Center for Astroparticle Physics, Universit\'{e} de Gen\`{e}ve, 24 quai Ernest Ansermet, CH-1211 Geneva, Switzerland\\
$^{16}$ Laborat\'orio Interinstitucional de e-Astronomia - LIneA, Rua Gal. Jos\'e Cristino 77, Rio de Janeiro, RJ - 20921-400, Brazil\\
$^{17}$ Institute of Astronomy, University of Cambridge, Madingley Road, Cambridge CB3 0HA, UK\\
$^{18}$ Kavli Institute for Cosmology, University of Cambridge, Madingley Road, Cambridge CB3 0HA, UK\\
$^{19}$ Department of Physics \& Astronomy, University College London, Gower Street, London, WC1E 6BT, UK\\
$^{20}$ Brookhaven National Laboratory, Bldg 510, Upton, NY 11973, USA\\
$^{21}$ Fermi National Accelerator Laboratory, P. O. Box 500, Batavia, IL 60510, USA\\
$^{22}$ Institute for Astronomy, University of Edinburgh, Edinburgh EH9 3HJ, UK\\
$^{23}$ Department of Physics and Electronics, Rhodes University, PO Box 94, Grahamstown, 6140, South Africa\\
$^{24}$ Institute of Cosmology \& Gravitation, University of Portsmouth, Portsmouth, PO1 3FX, UK\\
$^{25}$ CNRS, UMR 7095, Institut d'Astrophysique de Paris, F-75014, Paris, France\\
$^{26}$ Sorbonne Universit\'es, UPMC Univ Paris 06, UMR 7095, Institut d'Astrophysique de Paris, F-75014, Paris, France\\
$^{27}$ Observat\'orio Nacional, Rua Gal. Jos\'e Cristino 77, Rio de Janeiro, RJ - 20921-400, Brazil\\
$^{28}$ Department of Astronomy, University of Illinois at Urbana-Champaign, 1002 W. Green Street, Urbana, IL 61801, USA\\
$^{29}$ National Center for Supercomputing Applications, 1205 West Clark St., Urbana, IL 61801, USA\\
$^{30}$ Institut d'Estudis Espacials de Catalunya (IEEC), 08193 Barcelona, Spain\\
$^{31}$ Institute of Space Sciences (ICE, CSIC),  Campus UAB, Carrer de Can Magrans, s/n,  08193 Barcelona, Spain\\
$^{32}$ Centro de Investigaciones Energ\'eticas, Medioambientales y Tecnol\'ogicas (CIEMAT), Madrid, Spain\\
$^{33}$ Department of Astronomy, University of Michigan, Ann Arbor, MI 48109, USA\\
$^{34}$ Instituto de Fisica Teorica UAM/CSIC, Universidad Autonoma de Madrid, 28049 Madrid, Spain\\
$^{35}$ Department of Physics, ETH Zurich, Wolfgang-Pauli-Strasse 16, CH-8093 Zurich, Switzerland\\
$^{36}$ Santa Cruz Institute for Particle Physics, Santa Cruz, CA 95064, USA\\
$^{37}$ Harvard-Smithsonian Center for Astrophysics, Cambridge, MA 02138, USA\\
$^{38}$ Australian Astronomical Observatory, North Ryde, NSW 2113, Australia\\
$^{39}$ Departamento de F\'isica Matem\'atica, Instituto de F\'isica, Universidade de S\~ao Paulo, CP 66318, S\~ao Paulo, SP, 05314-970, Brazil\\
$^{40}$ Instituci\'o Catalana de Recerca i Estudis Avan\c{c}ats, E-08010 Barcelona, Spain\\
$^{41}$ Excellence Cluster Universe, Boltzmannstr.\ 2, 85748 Garching, Germany\\
$^{42}$ Faculty of Physics, Ludwig-Maximilians-Universit\"at, Scheinerstr. 1, 81679 Munich, Germany\\
$^{43}$ School of Physics and Astronomy, University of Southampton,  Southampton, SO17 1BJ, UK\\
$^{44}$ Brandeis University, Physics Department, 415 South Street, Waltham MA 02453\\
$^{45}$ Instituto de F\'isica Gleb Wataghin, Universidade Estadual de Campinas, 13083-859, Campinas, SP, Brazil\\
$^{46}$ Computer Science and Mathematics Division, Oak Ridge National Laboratory, Oak Ridge, TN 37831\\
$^{47}$ Cerro Tololo Inter-American Observatory, National Optical Astronomy Observatory, Casilla 603, La Serena, Chile\\
\bsp	
\label{lastpage}
\end{document}